\documentclass{iopart}

\usepackage{graphicx}

\graphicspath{{../fig/}}

\begin{document}
\title[Dendritic flux avalanches]{Dendritic flux avalanches in superconducting films of different thickness}

\author{J~I~Vestg{\aa}rden$^1$, Y~M~Galperin$^{1~2}$,  T~H~Johansen$^{1~3}$ }
\address{$^1$ Department of Physics, University of Oslo, PO Box 1048
Blindern, 0316 Oslo Norway}
\address{$^3$ Ioffe Physical Technical Institute, 26 Polytekhnicheskaya, St Petersburg 194021, Russian Federation}
\address{$^2$  Institute for Superconducting and Electronic Materials,
University of Wollongong, Northfields Avenue, Wollongong, NSW 2522,  Australia}

\ead{j.i.vestgarden@fys.uio.no}

\begin{abstract}
At low temperatures the critical state in superconducting films can be
unstable with respect to thermomagnetic dendritic avalanches. By numerical
simulations of disk-shaped superconductors, we consider how the
dynamics and morphology of the avalanches depend on the disk
thickness.  We find that as the disks get thicker, the jumps in
magnetic moment caused by the avalanches get larger and the threshold
magnetic field for the appearance of the first avalanche increases. At the same
time, the branches are straighter and the number of branches
decreases.  Comparison with theory suggests that strong spatial
disorder to some extent cancels the stabilizing effects of the
substrate kept at constant temperature.
\end{abstract}


\maketitle

\section{Introduction}
The critical state in type-II superconducting films subjected to transverse applied 
magnetic field or current can be  
susceptible to intermittent dynamics, where 
magnetic flux rushes in from the edges, 
forming large branching flux structures. The 
patterns remaining after such dendritic flux avalanches have been
imaged in many materials, e.g., 
Pb \cite{desorbo62},
Sn \cite{dolan73},
Nb \cite{duran95},
YBa$_2$Cu$_3$O$_{7-x}$ \cite{brull92},
MgB$_2$ \cite{johansen02},
Nb$_3$Sn \cite{rudnev03}, 
YNi$_2$B$_2$C \cite{wimbush04},
NbN \cite{rudnev05},
and a-MoGe \cite{motta13}.
Because the rapid motion of magnetic flux also implies a major redistribution 
of currents in the samples, the flux avalanches are associated 
with sudden drops in the magnetic moment values \cite{zhao02, choi04, colauto10, lee10-2}.

Dendritic flux avalanches are caused by
a thermomagnetic instability initiated when 
a temperature fluctuation facilitates uncontrolled penetration of magnetic flux and 
rise in temperature \cite{mints81}.
A model based on continuum electrodynamics and flow of heat 
has explained the phenomenon with great success, as numerical solutions 
has produced avalanche dynamics and patterns in striking resemblance with the experiments
\cite{aranson05,vestgarden11}, and linear stability analysis 
of the model has explained many features, such as the existence 
of a threshold temperature and magnetic field \cite{denisov06,albrecht07,vestgarden13-onset}
and a threshold electric field \cite{vestgarden13-metal}
for onset of avalanche activity.  
Experimentally it has been demonstrated that 
the threshold magnetic field increases with shrinking lateral size
\cite{denisov06,choi07} and simulations have shown that 
the properties of avalanches can be described by a small number 
of dimensionless parameters \cite{vestgarden13-diversity}.
The velocity of avalanches triggered by a laser-pulse has been
shown to be inversely proportional to the sample thickness \cite{bolz03}.
At the same time, it is not clear from previous works 
how the threshold magnetic field and avalanche morphology depend on the sample's thickness.

In this work, we consider how the properties of dendritic flux
avalanches depend on the sample thickness. We perform
numerical simulations for various thicknesses, but with otherwise
identical parameters, and consider the effect on the magnetic moment,
the threshold field, and the morphology of the dendritic flux
patterns.

\begin{figure}[b]
  \centering
  \includegraphics[width=7cm]{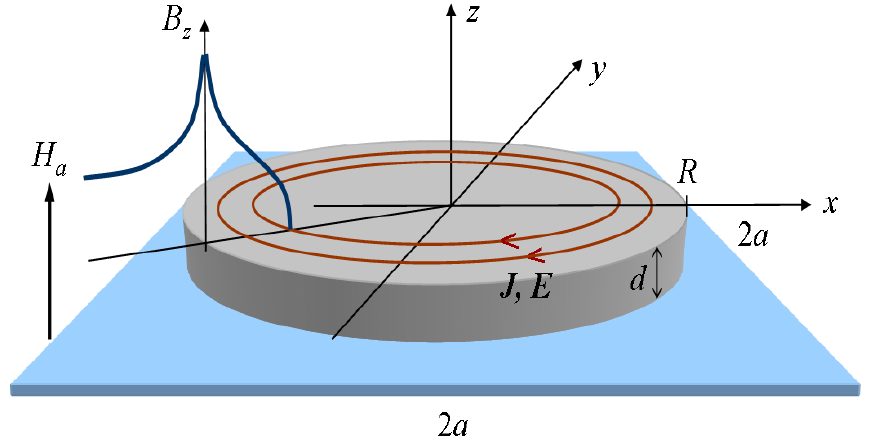} \
  \caption{
    \label{fig:sample}
    The sample is a disk of radius $R$ and thickness $d\ll R$.
    The magnetic field is applied transverse to the plane, causing 
    penetration of magnetic flux from the edges.
  }
\end{figure}

\section{Model}
\label{sec:model}
Let us consider a superconducting sample in gradually increasing 
transverse applied magnetic field $H_{\rm a}$, as depicted in figure \ref{fig:sample}. 
The film is shaped as a disk with radius $R$ and thickness $d$, where 
$d\ll R$.  In the flux creep regime, the resistivity is very non-linear as 
the sheet current $J$ approaches the sheet critical current, which we assume scales 
with sample thickness as $dj_c$, where $j_c$ is the critical current density. 
We use the conventional power law relation \cite{brandt97}
\begin{equation}
  \mathbf E =\rho \mathbf J/d,
  \quad
  \rho = \rho_0
  \left\{
  \begin{array}{ll}
    1, &T>T_{\rm c}~\mbox{or}~J>dj_{\rm c},\\
    \left(J/dj_{\rm c}\right)^{n-1}, &\mbox{otherwise},
  \end{array}
  \right.
  \label{power-law}
\end{equation}
where 
$n$ is the creep exponent and $\rho_0$ is the normal resistivity.
The parameters depend on local temperature $T$ as 
\begin{equation}
  j_{\rm c}=j_{\rm c0}\left(1-T/T_{\rm c}\right),\quad 
  n = n_1T_{\rm c}/T + n_0,
\end{equation}
for $T<T_{\rm c}$.

The fields must satisfy the Maxwell equations
\begin{equation}
  \nabla\times\mathbf H=\mathbf J\delta(z),\quad
  \nabla\cdot \mathbf B = 0,\quad
  \nabla\times\mathbf E = -\dot {\mathbf B},
\end{equation}
with $\mathbf B=\mu_0\mathbf H$ and $\nabla\cdot \mathbf J=0$. Here $\delta(z)$
is the Dirac delta function.

The local magnetization $g=g(x,y,t)$ is defined by
\begin{equation}
  \mathbf J=\nabla\times\hat zg=\nabla g\times \hat z
  .
\end{equation}
By performing the calculations using $g$ one assures that $\nabla\cdot \mathbf J=0$
holds at any time. The contour lines of $g$ are the current stream lines. 
Integrated over the sample area, $g$ gives the total magnetic moment,
\begin{equation}
  \label{def-moment}
  m   \hat z 
  = \frac{1}{2}\int \mathbf r\times\mathbf J  \rmd x\rmd y 
  = \hat z \int g \rmd x\rmd y 
  .
\end{equation}

From the Maxwell equations, one gets the time evolution of $g$ as \cite{vestgarden13-fftsim}
\begin{equation}
  \label{dotg}
  \dot g = {\mathcal F}^{-1}\left[\frac{2}{k}{\mathcal F}\left[\frac{1}{\mu_0}\dot B_z-\dot H_{\rm a}\right]\right]
  ,
\end{equation}
where $\mathcal F$ and $\mathcal F^{-1}$ are Fourier and inverse Fourier transforms, respectively,
and $k=|\mathbf k|$ is the wave-vector.

Inside the sample, $\dot B_z$ is
calculated from the Faraday law and the material law as 
\begin{equation}
  \label{dotBz}
  \dot B_z = \nabla\cdot (\rho\nabla g)/d
  .
\end{equation}
Outside the sample $\dot B_z$, is fixed by the condition $\dot g=0$, which is 
implemented using the iterative procedure described in \cite{vestgarden13-fftsim}.

The propagation of heat in the sample is governed by \cite{denisov06}
\begin{equation}
  \label{dotT}
  c\dot T = \kappa \nabla^2 T-h(T-T_0)/d+JE/d
  ,
\end{equation}
with specific heat $c$, thermal conductivity $\kappa$, coefficient of 
heat transfer to substrate $h$. The substrate is kept at constant temperature $T_0$.

The simulations are initiated with parameter values typical for 
films of MgB$_2$ \cite{denisov06, vestgarden11}:
$T_{\rm c}=39~$K,  $\rho_0$ = 7$\cdot 10^{-8}\Omega$m, 
$j_{\rm c0}= 1.2 \cdot 10^{11}$~A/m$^2$,
$n_1=20$, $n_0=-10$. 
The thermal parameters were
$\kappa=\left[0.17\right.$~kW/Km$\left.\right](T/T_{\rm c})^3$, 
$c = \left[35\right.$ ~kJ/Km$^3$$\left.\right](T/T_{\rm c}) ^3$, 
and 
$h = \left[200\right.$~kW/Km$^2$$\left.\right](T/T_{\rm c}) ^3$.
In all runs the applied magnetic field is driven with the same rate $\mu_0\dot H_{\rm a}=10~$T/s.

Let us write the equations on dimensionless form, in order to identify the effective 
parameters of the problem, and how they scale with $d$.
The dimensionless quantities are 
$\tilde g=g/Rdj_{\rm c0}$, $\tilde {\mathbf J}=\mathbf J/dj_{\rm c0}$, $\tilde j_{\rm c}=j_{\rm c}/j_{\rm c0}$, 
$\tilde H=H/dj_{\rm c0}$, $\tilde {\mathbf r} = \mathbf r/R$, 
$\tilde t = t \rho_0/\mu_0dR$, $\tilde E=E/\rho_0j_{\rm c0}$, $\tilde T =T/T_{\rm c}$,
$\rmd\tilde H_{\rm a}/\rmd\tilde t=\dot H_{\rm a}\mu_0R/j_{\rm c0}\rho_0$.
In these units, the material law becomes 
\begin{equation}
  \tilde {\mathbf E} =\tilde \rho \tilde{\mathbf J},
  \quad
  \tilde \rho = 
  \left\{
  \begin{array}{ll}
    1, &\tilde T>1~\mbox{or}~\tilde J>1-\tilde T,\\
    \left(\tilde J/(1-\tilde T)\right)^{n-1}, &\mbox{otherwise}.
  \end{array}
  \right.
\end{equation}
The Maxwell equations become
\begin{equation}
  \label{maxwell2}
  \tilde \nabla\times\tilde \mathbf H=\tilde \mathbf J\delta(\tilde z),\quad
  \tilde \nabla\cdot \tilde \mathbf H = 0,\quad
  \tilde \nabla\times \tilde \mathbf E = -{\rmd\tilde \mathbf H}/{\rmd \tilde t},
\end{equation}
and finally the heat propagation equation becomes 
\begin{equation}
  \label{dotT2}
  \frac{\rmd\tilde T}{\rmd\tilde t} 
  = \alpha\tilde \nabla^2\tilde T -
  \beta (\tilde T-\tilde T_0)+\gamma \tilde T^{-3}\tilde J\tilde E
  ,
\end{equation}
where 
\begin{equation}
  \label{alpha-beta-gamma}
  \begin{eqalign}
    \alpha\equiv \frac{d}{R}\frac{\mu_0}{\rho_n}\frac{\kappa}{c},\quad
    \beta \equiv R\frac{\mu_0}{\rho_n}\frac{h}{c},\quad
    \gamma \equiv Rd\frac{\mu_0}{c}\frac{j_{\rm c0}^2}{T_{\rm c}}
    .
  \end{eqalign}
\end{equation}
Here all parameters are evaluated at the critical temperature.

The state is found by discrete integration in time of (\ref{maxwell2}) and (\ref{dotT2}),
as described in \cite{vestgarden13-fftsim}.
The independent parameters of the dimensionless problem are:
$\alpha$, $\beta$, $\gamma$, $n_1$, and $\rmd \tilde H/\rmd \tilde t$.
Since both $\alpha$ and $\gamma$ depend on $d$, it is not possible 
to predict the thickness-dependency of the results just 
from inspection of the dimensionless parameters.

\begin{figure}[t]
  \centering
  \includegraphics[width=7cm]{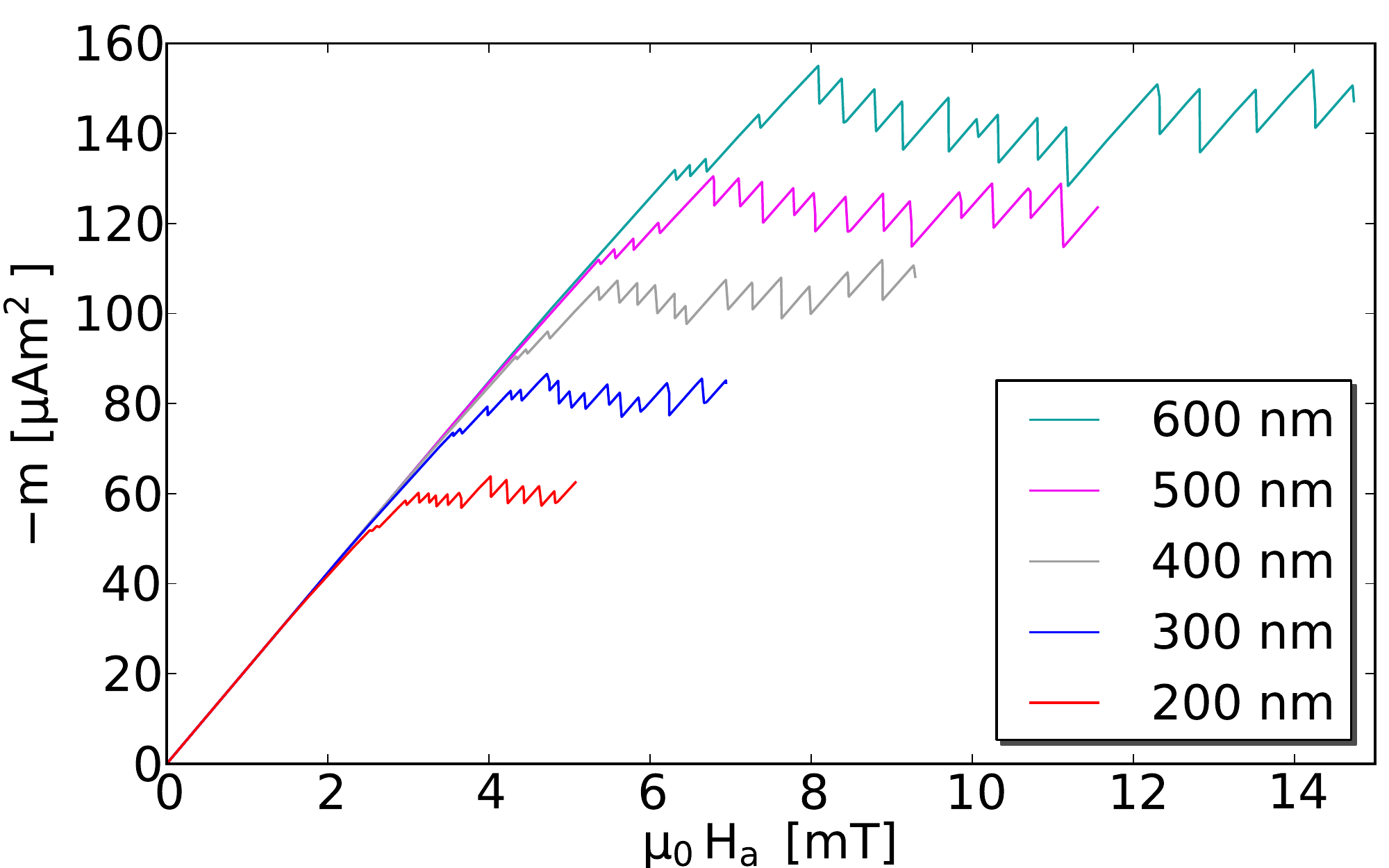} \\
  \includegraphics[width=7cm]{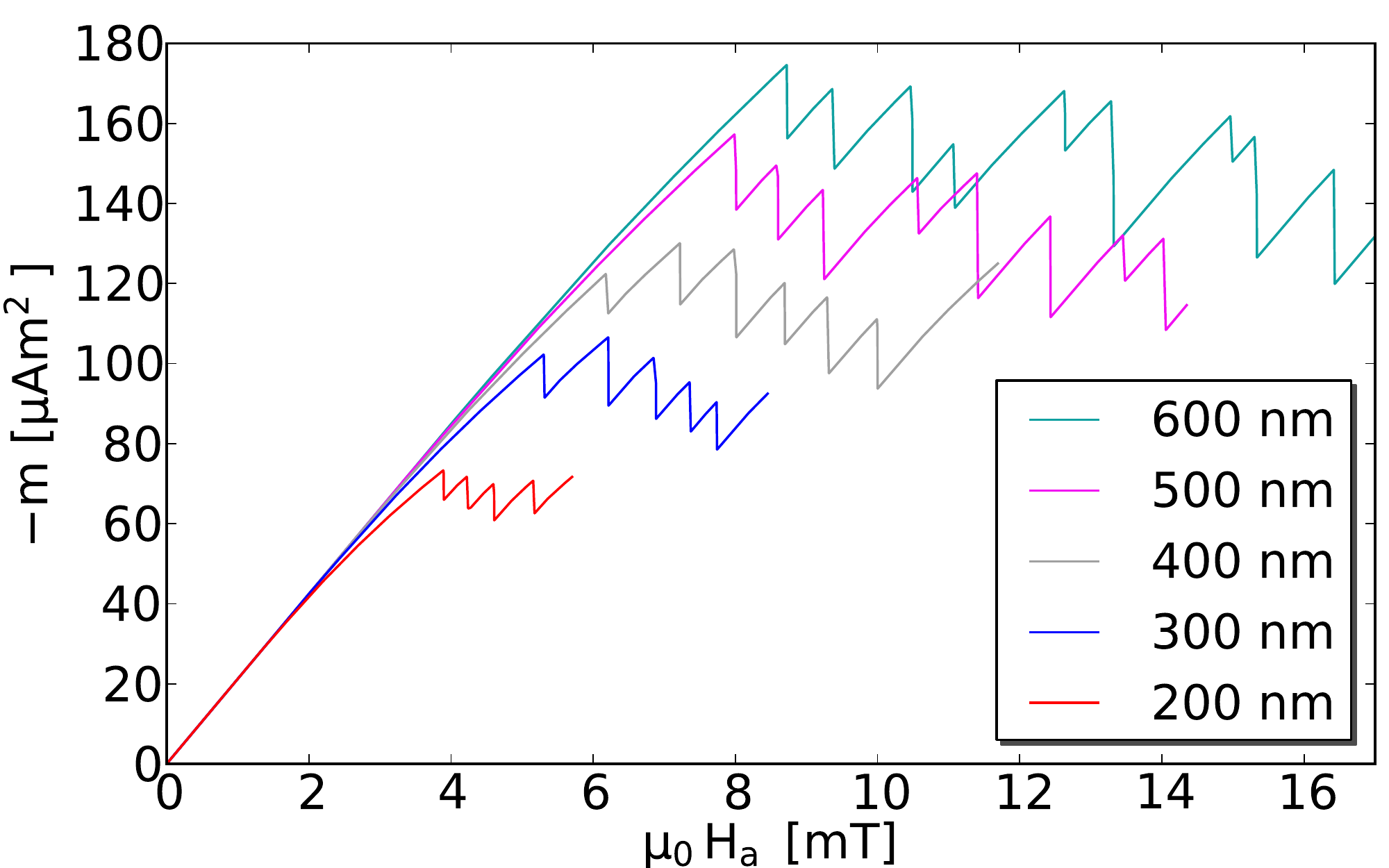} 
  \caption{The magnetic moment as function of applied field, with $d=0.2 - 0.6 \mu$m,    
    at $T=0.15$ (top) and $0.2T_{\rm c}$ (bottom). Each jump in the curves 
    corresponds to a dendritic flux avalanche.
    \label{fig:m1}
  }
\end{figure}

The disk-shaped specimen was embedded in a $a\times a$ square, 
with $a=1.2R$, where the extra space was used 
for implementation of the boundary conditions. The embedding 
square was discretized on a $512\times 512$ equidistant grid.
Spatial disorder was included into the 
formalism by randomly changing the value of $dj_{\rm c0}$ 
in each grid point by $\pm 5\%$. All runs have the same realization 
of spatial disorder.

\section{Results}
\label{sec:results}
Starting from zero-field-cooled state, an applied magnetic field is
gradually increased with constant rate $\dot H_{\rm a}$. Then a 
critical state is formed from the edges, with $J=dj_{\rm c}$ and
nonuniform $B_z$, which is highly peaked at the edge and falls to zero
at the flux front.  Inside the flux front the superconductor is in the flux-free
Meissner state, where $B_z=0$, but $\mathbf J\neq 0$ as a consequence
of the nonlocal electrodynamics.

Figure~\ref{fig:m1} shows the magnetic moment $m$ as a function of
applied field, extracted from the simulations during the field-ramp
using (\ref{def-moment}). Each curve corresponds to a different
thickness $d=0.2-0.6\mu$m, and the two panels are at substrate
temperatures $T_0=0.15$ and $0.2T_{\rm c}$.  Qualitatively, all curves have
the same behaviour.  Initially, the magnetic moment increases
smoothly as predicted by the critical state model, 
until the first avalanches appears as a jump in the curve 
at the threshold field $H_{\rm th}$. 
At $T_0=0.15T_{\rm c}$ the first jumps are quite small, but except from that,
most jumps are of comparable sizes, and each one of them is clearly
visible.  The figure shows that the size of jumps in magnetization are
larger for increasing $d$.  Unlike in bulks, where the thermomagnetic
instability typically causes a global breakdown in superconductivity, with
consequent magnetization drop to zero \cite{kim63,zhou06}, the
magnetization values of the figure are always nonzero, fluctuating
around a more or less constant value, c.f. \cite{choi07,colauto08}.

\begin{figure}[b]
  \centering
  \includegraphics[width=6cm]{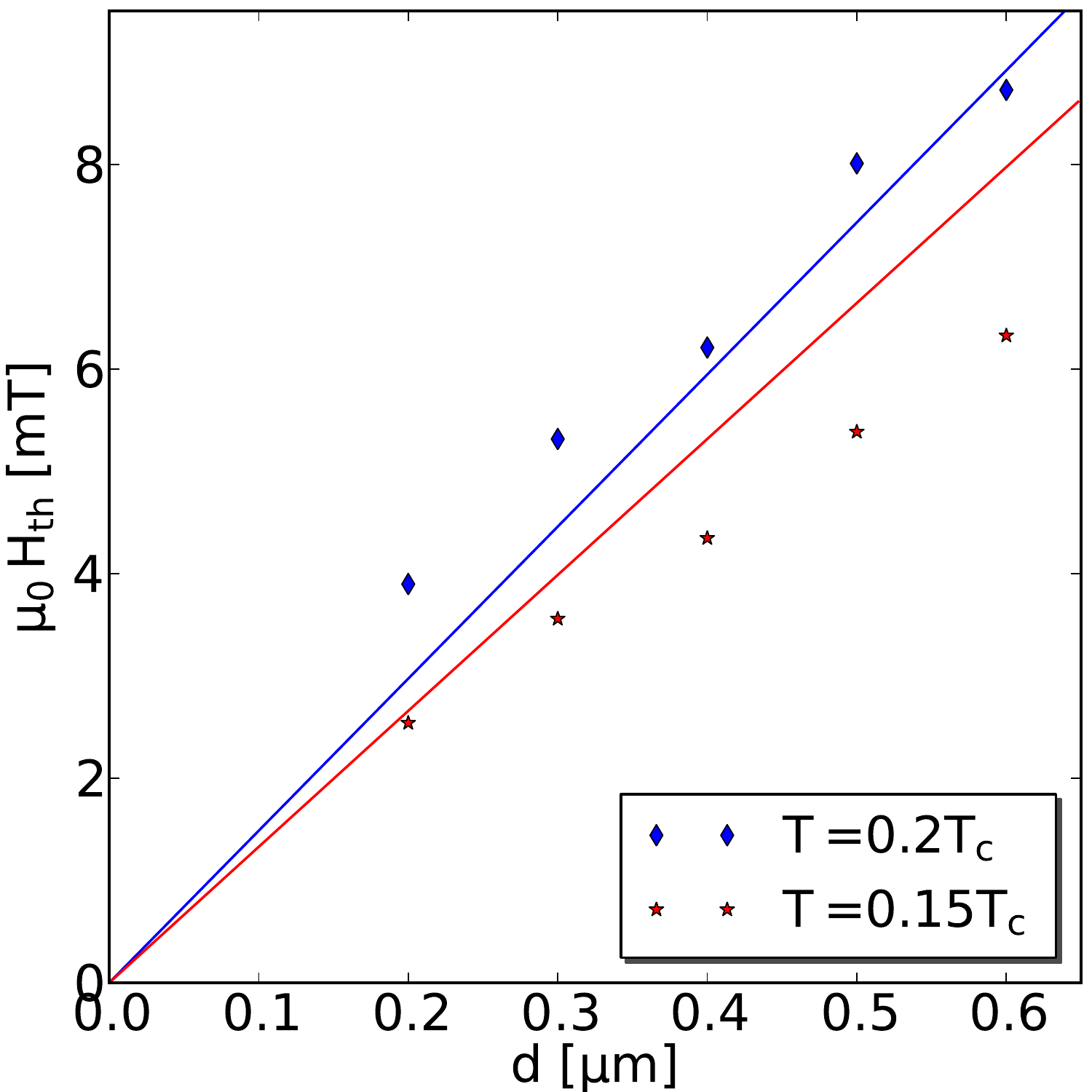} \
  \caption{
    \label{fig:Hth}
    The threshold field as a function of sample thickness.
    The points are extracted from the simulations, the lines 
    are plots of the analytical prediction, $H_{\rm th}$.
  }
\end{figure}

Let us see to what extent the thresholds for onset of
instability in the simulations agree with the criteria from linear stability analysis.  In 
disks, the magnetic flux penetration depth as a function of
applied field is \cite{mikheenko93} 
\begin{equation}
  l = R-R/\cosh(H_{\rm a}/H_{\rm c})
  ,
\end{equation}
where $H_{\rm c}=dj_{\rm c}/2$. Since there is only a numerical factor difference
compared to strips, where $H_{\rm c}=dj_{\rm c}/\pi$ \cite{brandt93-epl}, we can 
reuse the formulas for the threshold field previously derived for
strips, only with a change of numerical constants.  By assuming
that the most unstable mode is spatially constant one gets \cite{vestgarden13-onset}
\begin{equation}
  \label{Hth}
  H_{\rm th} = \frac{dj_{\rm c}}{2}
  \left[
    \frac{\pi^2\kappa T^*}{nR^3j_{\rm c}\mu_0\dot H_{\rm a}}
    \right]^\frac{1}{5}
  ,
\end{equation}
where $1/T^*\equiv \left|\partial \log j_{\rm c}/\partial T\right|$.

\begin{figure*}[t]
  \centering
  \includegraphics[width=13cm]{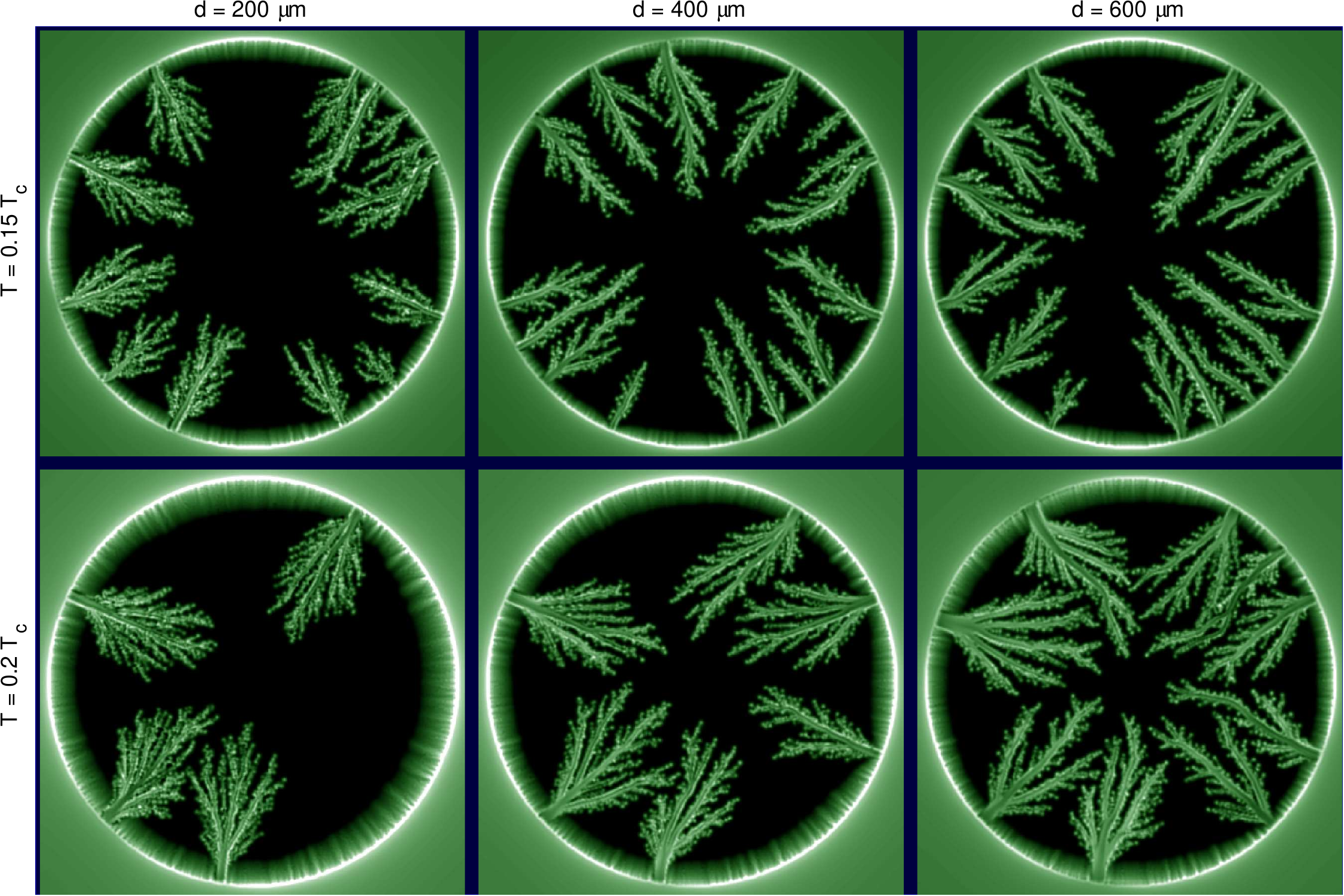} \
  \caption{
    The $B_z$ distributions for $d=0.2$, 0.4 and 0.6~$\mu$m and $T_0=0.15T_{\rm c}$ (top)
    $T_0=0.2T_{\rm c}$ (bottom). The applied field is $H_{\rm a}=0.15dj_{\rm c0}$ (top) and 
    and $H_{\rm a}=0.19dj_{\rm c0}$ (bottom).
    \label{fig:Bz}
  }
\end{figure*}

Figure \ref{fig:Hth} shows the threshold field as a function of sample
thickness, with $T_0=0.15$ and 0.2$T_{\rm c}$.  The discrete points in
the figure are extracted from the simulations (the magnetic moment
curves of figure~\ref{fig:m1}). The figure shows that $H_{\rm th}$
increases with $d$, and that the increase is close to linear. At the
same time $H_{\rm th}$ increases with $T_0$, as expected from previous
theory and experiments \cite{denisov06, vestgarden13-onset}.  To make
an interpretation of the simulation results, figure \ref{fig:Hth} also
plots the analytical $H_{\rm th}$, equation (\ref{Hth}), for the two
temperatures. We see that the analytical predictions are matching the
simulation results quite well. This indicates that the instability is
prevented mainly by the lateral heat diffusion, which is the only
mechanism included in (\ref{Hth}). In particular, the $H_{\rm
  th}\propto d$ dependency seen both in the analytical curve and the
numerical results is typical for avalanches being prevented by the
lateral heat diffusion. A deviation from proportionality would on the
contrary indicate the presence of surface effects, such as the heat
removal to the substrate. The absence of such effects and comparison with
previous works \cite{vestgarden13-onset} indicate that the
stabilizing effect of the substrate is being neutralized by the presence of
the spatial disorder used in the simulation. 

Figure \ref{fig:Bz} shows the flux distributions for $d=0.2$, 0.4 and 0.6~$\mu$m,
and $T_0=0.15$ and $0.2T_{\rm c}$. All panels are at different times, as the applied fields 
are $H_{\rm a}=0.15dj_{\rm c0}$ and  $0.19dj_{\rm c0}$ for $T_0=0.15T_{\rm c}$ and  $0.2T_{\rm c}$, respectively.
We see that the avalanches are significantly 
larger at the highest temperature, and that they have got more 
branches and more complex morphology.
Also the sample thickness seems to affect the morphology of the avalanches,
in the sense that increasing $d$ gives straighter avalanches, with fewer 
and thicker branches. This dependency is partly a consequence of the 
changing depth of penetration prior to the avalanches, partly a consequence of 
changing heat removal to substrate during the avalanches.

Because all runs are with identical realization of the spatial disorder,
there is some overlap between the nucleation spots of the avalanches.
This implies that the avalanches are not nucleated by uniform instabilities,
which is the case for the thermomagnetic instability in spatially 
uniform samples \cite{vestgarden13-onset}, but instead they appear
on places selected by an interplay between the randomly distributed 
disorder and fluctuations in the electric field values.

\section{Summary}
\label{sec:summary}
The critical state in superconducting films can, at low temperatures, be unstable 
with respect to thermomagnetic avalanches. We have considered 
how the properties of avalanches  
depend on the sample's thickness by performing numerical simulations 
on disks of various thicknesses, but otherwise identical parameters.

We have shown that a ticker sample gives a larger jump in the magnetic moment 
and higher threshold field for the appearance of the first avalanche.
At the same time the branches get straighter and 
the number of branches decreases. 
The threshold field in the simulation grows linearly with sample thickness
and matches the theoretical prediction where the only mechanism included was the lateral heat transport.
Due to the strong spatial disorder, the substrate kept 
at constant temperature has only minor effect on the stability.

\ack
This work was financially supported by the Research Council of Norway.

\bibliographystyle{unsrt}

\bibliography{superconductor}

\end{document}